\documentclass{jpsj3}

\usepackage{txfonts}

\usepackage{siunitx}

\usepackage{algpseudocode}
\usepackage{algorithm}

\usepackage{graphicx}

\usepackage{amsmath}
\usepackage{mathtools}
\usepackage{bm}

\newcommand{\argmin}{\mathop{\rm arg~min}\limits}
\DeclareMathOperator*{\extr}{extr}
\usepackage{comment}

\title{Phase transition in binary compressed sensing based on $L_{1}$-norm minimization}

\author{Mikiya Doi$^1$ and Masayuki Ohzeki$^{1,2,3}$}
\inst{Graduate School of Information Sciences, Tohoku University, Miyagi 980-8564, Japan$^1$ \\
Department of Physics, Tokyo Institute of Technology, Tokyo, 152-8551, Japan$^2$ \\
Sigma-i Co., Ltd., Tokyo, 108-0075, Japan$^3$
} 

\abst{
Compressed sensing is a signal processing scheme that reconstructs high-dimensional sparse signals from a limited number of observations.
In recent years, various problems involving signals with a finite number of discrete values have been attracting attention in the field of compressed sensing.
In particular, binary compressed sensing, which restricts signal elements to binary values $\{0, 1\}$, is the most fundamental and straightforward analysis subject in such problem settings.
We evaluate the typical performance of noiseless binary compressed sensing based on $L_{1}$-norm minimization using the replica method, a statistical mechanical approach.
We analyze a general setting where the elements of the observation matrix follow a Gaussian distribution, including a non-zero mean.
We demonstrate that the biased observation matrix indicates more reconstruction success conditions in binary compressed sensing.
Our results are consistent with the outcomes of several prior studies.
}

\begin{document}
\maketitle
\renewcommand{\algorithmicrequire}{\textbf{Input:}}
\renewcommand{\algorithmicensure}{\textbf{Output:}}

\section{Introduction.}
Compressed sensing is the scheme that reconstructs signals from linear observations when the number of observations is smaller than the dimension of the signal to be reconstructed \cite{Donoho2006, Candes2008-2}.
When the number of observations is insufficient, the signal reconstruction problem is mathematically formulated as an underdetermined system, where the solution is not uniquely determined.
However, it is possible to reconstruct in some cases if the signal to be reconstructed is sparse, meaning it contains zero components with a finite rate.
Theories and techniques of compressed sensing have been developed by research in multiple disciplines such as mathematics \cite{Donoho2005, candes2005DecodingIEEETrans.Inform.Theory, candes2006NearOptimalIEEETrans.Inf.Theory, Candes2008}, signal processing \cite{Angelosante2009, Ye2019}, and statistical physics \cite{kabashima2009TypicalJ.Stat.Mech., Krzakala2012}.

In recent years, within the field of compressed sensing, problem settings that involve signals composed of a finite number of discrete values are gaining attention in contexts such as digital communications \cite{Sparrer2015, Fosson2019, Fukshansky2019}, quantization systems \cite{Abhik2013}, group testing \cite{atia2012BooleanIEEETrans.Inform.Theory, malioutov2012Boolean2012IEEEInt.Conf.Acoust.SpeechSignalProcess.ICASSP, feige2020Quantitative} detection of irregular pairs \cite{Yamanaka2015}, decoding \cite{ide2022}, estimation of hidden interactions \cite{Ohzeki2015, Ezaki2017, Takahashi2018, Ohzeki2018}, visualization of dynamical behavior \cite{Sogabe2020, Yamamoto2023} and even quantum many-body physics \cite{Otsuki2017, Shinaoka2017, Shinaoka2018, Ohzeki2018jpcs, Yoshimi2019, Otsuki2020}. 
In particular, binary compressed sensing, which restricts signal elements to binary values $\{0, 1\}$, is the most common and straightforward setting \cite{amini2011DeterministicIEEETrans.Inform.Theory, 6415872, keiper2017CompressedLinearAlgebraanditsApplications}.
Noiseless binary compressed sensing is reduced to finding sparse solutions to the following linear equations.
\begin{align}
    \bm{y} \in \mathbb{R}^{M}, \,\,\, A \in \mathbb{R}^{M \times N}, \,\,\, \bm{y} = A \bm{x},
    \label{eq:linear_equation}
\end{align}
where $A$ is the observation matrix$(M<N)$, and the observed signal $\bm{y}$ is generated from the unknown binary original signal $\bm{x}^{0} \in \{0, 1\}^{N}$ by $\bm{y} = A \bm{x}^{0}$. 
Compressed sensing is often formulated as a problem of optimizing the regularization functions of the signals to be estimated. \\
A natural approach is solving the $L_{0}$-norm minimization problem as the discrete optimization problem:
\begin{align}
    \hat{\bm{x}} = \argmin_{\bm{x} \in \{0, 1\}^{N}}{\| \bm{x}\|_{0}}
    \,\,\,\,\,\, \text{subject to} \,\,\,\, \bm{y} = A \bm{x},
    \label{eq:L0_minimization}
\end{align}
where $\|\bm{x}\|_{0}$ is the number of non-zero components in $\bm{x}$.
However, the $L_{0}$-norm minimization approach is known to be impractical from the perspective of computational complexity \cite{Ayanzadeh2019}.
Hence, one of the major approaches is solving the $L_{1}$-norm minimization problem as the relaxed problem, defined as follows:
\begin{align}
    \hat{\bm{x}} = \argmin_{\bm{x} \in \{0 \leq x_{i} \leq 1\}^{N}}{\| \bm{x}\|_{1}}
    \,\,\,\,\,\, \text{subject to} \,\,\,\, \bm{y} = A \bm{x}.
    \label{eq:L1_minimization}
\end{align}
Since the optimization problem shown as Eq.(\ref{eq:L1_minimization}) is convex, it can be solved effectively with linear programming and convex optimization techniques.
Recently, a theoretical typical performance of noiseless binary compressed sensing based on $L_{1}$-norm minimization shown in Eq.(\ref{eq:L1_minimization}) has been demonstrated \cite{keiper2017CompressedLinearAlgebraanditsApplications, flinth2019RecoveryIEEETrans.Inform.Theory}.

Numerous studies have shown that the typical performance of compressed sensing depends on the choice of regularization function \cite{fan2001VariableJ.Am.Stat.Assoc., zhang2010NearlyAnn.Stat., sakata2018ApproximateJ.Stat.Mech.a, sakata2021PerfectJ.Stat.Mech., nagano2023PhasePhys.Rev.E} and construction of the observation matrix \cite{tanaka2018Performance2018IEEEInt.Symp.Inf.TheoryISIT, Naidu2016}.
Regarding the latter, it has been shown that an observation matrix with biased elements can estimate the original signal with fewer observations when there is an imbalance of positive and negative values in the non-zero components of the original signal \cite{tanaka2018Performance2018IEEEInt.Symp.Inf.TheoryISIT}.
In addition, similar remarks have been made in a different study that analyzed binary compressed sensing \cite{flinth2019RecoveryIEEETrans.Inform.Theory}.
Since most of the existing analytical techniques rely on the assumption of zero mean elements, analyses targeting observation matrices with non-zero means have not been widely performed.
It is also likely related to the unstable behavior of the approximate message passing (AMP) algorithm \cite{donoho2009MessagepassingProc.Natl.Acad.Sci.U.S.A.}, an effective algorithm in compressed sensing when using observation matrices with non-zero means \cite{Caltagirone2014}.

The contribution of this paper is demonstrating the typical performance and signal recovery phase transition in the binary compressed sensing based on $L_{1}$-norm minimization by replica method, the statistical mechanical approach.
Specifically, our analysis provides a comprehensive typical evaluation for both zero mean and non-zero mean elements of the observation matrix, as shown in prior studies \cite{keiper2017CompressedLinearAlgebraanditsApplications, flinth2019RecoveryIEEETrans.Inform.Theory}.
In addition, we verify our analytical results through numerical simulations using Gurobi Optimizer, a commercial mathematical optimization solver \cite{gurobi}.

The remainder of this paper is organized as follows.
In Section 2, we briefly introduce our problem settings.
Section 3 presents the analysis using the replica method and shows the obtained theoretical phase boundaries.
In Section 4, we compare the results of the numerical simulations with our theoretical analysis presented in the previous section.
In the final Section 5, we summarize our results and discuss future studies. We provide details of the analysis in Appendix.
\section{Problem Settings}
We assume that each element $x_{i}^{0}$ of the binary original signal $\bm{x}^{0}$ to be reconstructed, and each element $A_{ij}$ of the observation matrix $A$, are generated from the following probability distributions. 
\begin{align}
    & x_{i}^{0} \sim \rho \delta(x_{i}^{0} - 1) + (1 - \rho)\delta(x_{i}^{0}) \label{eq:original_signal_i}\\
    & A_{ij} \sim \mathcal{N} \left(\frac{\gamma}{N}, \frac{1}{N} \right), \label{eq:observation_matrix_element}
\end{align}
where $\rho$ represents the non-zero elements rate in the binary original signal.
In addition, to reiterate, the observed signal $\bm{y}$ is assumed to be generated as $\bm{y} = A \bm{x}^{0}$.
\section{Replica Analysis}
The approach based on $L_{1}$-norm minimization shown in Eq.(\ref{eq:L1_minimization}) is derived from the Bayesian framework.
In noiseless binary compressed sensing, the likelihood $p(\bm{y}|A, \bm{x})$ is given as follows.
\begin{align}
    p(\bm{y}|A, \bm{x}) = \delta(\bm{y} - A \bm{x}) .\label{eq:likelihood}
\end{align}
We assume the Laplace prior for $\bm{x}$ as
\begin{align}
    p(\bm{x}) \propto \exp(-\beta \|\bm{x} \|_{1}), \,\,\,\,\, \beta \geq 0. \label{eq:prior}
\end{align}
Then the posterior distribution of $\bm{x}$ given $A$ and $\bm{y}$ is represented as
\begin{align}
    p(\bm{x}|A, \bm{y}) = \frac{p(A, \bm{y}| \bm{x})p(\bm{x})}{Z_{\beta}(A, \bm{y})}
    = 
    \frac{\delta(\bm{y} - A \bm{x})\exp(-\beta \|\bm{x} \|_{1})}
    {\int_{0}^{1} d\bm{x}\, \delta(\bm{y} - A \bm{x})\exp(-\beta \|\bm{x} \|_{1})}, \label{eq:posterior}
\end{align}
where $Z_{\beta}(A, \bm{y})$ is a normalized constant.
By considering the limit as $\beta \rightarrow \infty$, it can be confirmed that $L_{1}$-norm minimization corresponds to \textit{maximum a posteriori} (MAP) estimation.
From a perspective of statistical physics, the regularization function, the $L_{1}$-norm, can be viewed as a Hamiltonian $H(\bm{x}) = \|\bm{x}\|_{1}$, and the limit as $\beta \rightarrow \infty$ corresponds to zero-temperature statistical mechanics.
In addition, the normalization constant $Z_{\beta}(A, \bm{y})$ corresponds to the partition function.
The logarithm of the partition function $Z_{\beta}(A, \bm{y})$ is used when calculating the free energy, an important physical quantity.
Note that our subsequent analysis based on statistical physics is performed with the compression ratio $\alpha = M/N$ and the non-zero elements rate in the original signal $\rho$ fixed, considering the limit as $N \rightarrow \infty$.
Following the standard prescription of the replica analysis \cite{Nishimori2001}, the free energy density $f$ can be evaluated as follows, assuming self-averaging property:
\begin{align}
    f
    = 
    - \lim_{\beta \rightarrow \infty} \lim_{N \rightarrow \infty} \frac{1}{\beta N}
        \left[ 
            \ln Z_{\beta}(A, \bm{y})
        \right]_{A, \bm{x}^{0}} 
    =
    - \lim_{\beta \rightarrow \infty}
      \lim_{n \rightarrow 0}
      \frac{\partial}{\partial n}
      \lim_{N \rightarrow \infty} \frac{1}{\beta N}    
            \ln \left[ Z_{\beta}^{n}(A, \bm{y})
      \right]_{A, \bm{x}^{0}} ,
      \label{eq:free_energy_density}
\end{align}
where $[]_{A, \bm{x}^{0}}$ represents the configurational average with respect to $A$ and $\bm{x}^{0}$.
Assuming $n$ is a positive integer, we can express the expectation of $Z_{\beta}^{n}(A, \bm{y})$ as a replicated system.
We assess $Z_{\beta}^{n}(A, \bm{y})$ in Eq.(\ref{eq:free_energy_density}) for $n \in \mathbb{Z}_{+}$, then the free energy density is evaluated by performing an analytic continuation for $n \in \mathbb{R}$. The following equation explicitly expresses the replicated system:
\begin{align}
    \left[Z_{\beta}^{n}(A, \bm{y})\right]_{A, \bm{x}^{0}}
    =
    \left[
        \prod_{a=1}^{n}
        \int_{0}^{1} d \bm{x}^{a} 
        \prod_{\mu=1}^{M}
        \delta(y_{\mu} - \bm{a}_{\mu}^{T} \bm{x}^{a}) \exp(-\beta \|\bm{x}^{a}\|_{1})
    \right]_{A, \bm{x}^{0}} ,
    \label{eq:replicated_system}
\end{align}
where $\bm{x}^{a}$ represents the vector for the $a$-th replica, $y_{\mu}$ represents the $\mu$-th element of the observed signal $\bm{y}$, and $\bm{a}_{\mu}^{T}$ stands for the $\mu$-th row of the observation matrix $A$.
To proceed with the analysis of the equation, we define the following quantity:
\begin{align}
    u_{\mu}^{a} = y_{\mu} - \bm{a}_{\mu}^{T} \bm{x}^{a} 
                = \bm{a}_{\mu}^{T}(\bm{x}^{0} - \bm{x}^{a}) .
                \label{eq:u_transformed}
\end{align}
Since each element of the observation matrix follows the Gaussian distribution with mean $\gamma/N$ and covariance $1/N$, the expected value and covariance for $u_{\mu}^{a}$ are as follows:  
\begin{align*}
    & \mathbb{E}\left[u_{\mu}^{a} \right]_{A} 
    = 
    \frac{\gamma}{N} \sum_{i=1}^{N} x_{i}^{0} 
    - \frac{\gamma}{N}\sum_{i=1}^{N}x_{i}^{a}
    \nonumber \\
    & \text{Cov} \left[u_{\mu}^{a}, u_{\mu}^{b} \right]_{A}
    =
    \frac{1}{N}{\bm{x}^{0}}^{T} \bm{x}^{0} 
    -
    \frac{2}{N}{\bm{x}^{0}}^{T} \bm{x}^{a} 
    +
    \frac{1}{N}{\bm{x}^{a}}^{T} \bm{x}^{b} .
\end{align*}
We define the order parameters under the replica symmetric (RS) ansatz.
\begin{align}
    & \rho = \frac{1}{N}\sum_{i=1}^{N}x_{i}^{0} = \frac{1}{N}{\bm{x}^{0}}^{T} \bm{x}^{0} \,\,\,\,\, (\because x_{i}^{0} \in \{0, 1\}) \label{eq:order_params_rho}\\
    & p = \frac{1}{N} \sum_{i=1}^{N}x_{i}^{a} \,\,\,\,\, (a = 1 \sim n) \label{eq:order_params_p} \\
    & m = \frac{1}{N}{\bm{x}^{0}}^{T} \bm{x}^{a} \,\,\,\,\, (a = 1 \sim n) \label{eq:order_params_m} \\
    & Q = \frac{1}{N}{\bm{x}^{a}}^{T} \bm{x}^{a} \,\,\,\,\, (a = 1 \sim n) \label{eq:order_params_Q} \\
    & q = \frac{1}{N}{\bm{x}^{a}}^{T} \bm{x}^{b} \,\,\,\,\, (a,b = 1 \sim n, \, a \neq b). \label{eq:order_params_q}
\end{align}
With the order parameters defined Eqs.(\ref{eq:order_params_rho})-(\ref{eq:order_params_q}), and standard Gaussian random variables $\xi_{a}$ and $z$, we can parameterize $u_{\mu}^{a}$ in Eq.(\ref{eq:u_transformed}) as follows:
\begin{align}
    u_{\mu}^{a} = \gamma(\rho - p ) + \sqrt{Q - q} \, \xi_{a} + \sqrt{\rho - 2m + q} \, z.
\end{align}
We provide the detailed analysis in Appendix, the free energy density $f$ in the limit as $\beta \rightarrow \infty$, under the replica symmetric ansatz, is as follows:
\begin{align}
    f = \extr_{Q, m, \chi, p, \tilde{Q}, \tilde{m}, \tilde{\chi}, \tilde{p}}
    \left\{
    - \frac{1}{2}Q\tilde{Q} + m\tilde{m} + \frac{1}{2}\chi \tilde{\chi} + p \tilde{p} \right. 
    + \frac{\alpha\left[(\rho - 2m + Q) + \gamma^{2} (\rho - p)^{2}\right]}{2\chi} 
    \nonumber \\
    + \rho\int Dt \, \Psi (t; \tilde{Q}, \tilde{m}, \tilde{\chi}, \tilde{p}) \left. 
    + (1 - \rho)\int Dt \, \Psi (t; \tilde{Q}, 0, \tilde{\chi}, \tilde{p})
    \right\} ,
    \label{eq:extr_free_energy_density}
\end{align}
where $\int Dt = \int \frac{dt}{\sqrt{2 \pi}} \exp \left(- \frac{1}{2}t^{2} \right)$, and we define $\Psi (t; \tilde{Q}, \tilde{m}, \tilde{\chi}, \tilde{p})$ as follows:
\begin{align*}
    \Psi (t; \tilde{Q}, \tilde{m}, \tilde{\chi}, \tilde{p}) 
    =
    \min_{\{0 \leq x \leq 1\}} 
    \left\{
        \frac{1}{2} \tilde{Q}x^{2} 
        - \left(\tilde{m} + \sqrt{\tilde{\chi}} t + \tilde{p} \right)x
        + \|x\|_{1}
    \right\} .
\end{align*}
The extremization of Eq.(\ref{eq:extr_free_energy_density}) yields the following saddle-point equations:
\begin{align}
    \tilde{Q} & = \frac{\alpha}{\chi} \label{eq:extr_Qtilde}\\
    \tilde{m} & = \frac{\alpha}{\chi} \label{eq:extr_mtilde}\\
    \tilde{\chi} & = \frac{\alpha\left[(\rho - 2m + Q) + \gamma^{2} (\rho - p)^{2}\right]}{\chi^{2}}\label{eq:extr_chitilde}\\
    \tilde{p} & = \frac{\alpha \gamma^{2}(\rho - p)}{\chi} \label{eq:extr_ptilde}\\
    Q & = \rho A(\tilde{Q}, \tilde{m}, \tilde{\chi}, \tilde{p}) + (1-\rho)A(\tilde{Q}, 0, \tilde{\chi}, \tilde{p}) \label{eq:extr_Q}\\
    m & = \rho B(\tilde{Q}, \tilde{m}, \tilde{\chi}, \tilde{p}) \label{eq:extr_m}\\ 
    \chi & = \rho C(\tilde{Q}, \tilde{m}, \tilde{\chi}, \tilde{p}) + (1- \rho) C(\tilde{Q}, 0, \tilde{\chi}, \tilde{p}) \label{eq:extr_chi}\\
    p & = \rho B(\tilde{Q}, \tilde{m}, \tilde{\chi}, \tilde{p}) + (1 - \rho) B(\tilde{Q}, 0, \tilde{\chi}, \tilde{p}).
    \label{eq:extr_p}
\end{align}
Here, we define
{\small
\begin{align*}
    A(\tilde{Q}, \tilde{m}, \tilde{\chi}, \tilde{p}) & = H \left(\frac{\tilde{Q} - \tilde{m} - \tilde{p} + 1}{\sqrt{\tilde{\chi}}} \right) - \frac{\tilde{Q} + \tilde{m} + \tilde{p} - 1} {\tilde{Q}^{2}}\sqrt{\frac{\tilde{\chi}}{2\pi}} \exp \left( - \frac{(\tilde{Q} - \tilde{m} - \tilde{p} + 1)^{2}}{2\tilde{\chi}}\right) 
    - \frac{1 - \tilde{m} - \tilde{p}}{\tilde{Q}^{2}}\sqrt{\frac{\tilde{\chi}}{2\pi}} \exp \left( - \frac{(1 - \tilde{m} - \tilde{p})^{2}}{2\tilde{\chi}}\right)
    \\
    & + \frac{1}{\tilde{Q}^{2}}(\tilde{\chi} + (1 -\tilde{m} - \tilde{p})^{2}) 
    \left(H \left(\frac{1 -\tilde{m} - \tilde{p}}{\sqrt{\tilde{\chi}}} \right) - H \left(\frac{\tilde{Q} - \tilde{m} - \tilde{p} + 1}{\sqrt{\tilde{\chi}}} \right)\right) 
    \\
    B(\tilde{Q}, \tilde{m}, \tilde{\chi}, \tilde{p}) & = H \left(\frac{\tilde{Q} - \tilde{m} - \tilde{p} + 1}{\sqrt{\tilde{\chi}}} \right) - \frac{1}{\tilde{Q}}\sqrt{\frac{\tilde{\chi}}{2\pi}}
    \left(\exp \left( - \frac{(\tilde{Q} - \tilde{m} -\tilde{p} + 1)^{2}}{2\tilde{\chi}}\right) 
    - \exp \left( - \frac{(1 - \tilde{m} - \tilde{p})^{2}}{2\tilde{\chi}}\right) \right) 
    \\
    & - \frac{1 - \tilde{m} - \tilde{p}}{\tilde{Q}}
    \left(H \left(\frac{1 -\tilde{m} - \tilde{p}}{\sqrt{\tilde{\chi}}} \right) - H \left(\frac{\tilde{Q} - \tilde{m} -\tilde{p} + 1}{\sqrt{\tilde{\chi}}} \right)\right)
    \\
    C(\tilde{Q}, \tilde{m}, \tilde{\chi}, \tilde{p}) & = \frac{1}{\tilde{Q}} \left(H \left(\frac{1 -\tilde{m} - \tilde{p}}{\sqrt{\tilde{\chi}}} \right) - H \left(\frac{\tilde{Q} - \tilde{m} - \tilde{p} + 1}{\sqrt{\tilde{\chi}}} \right)\right),
\end{align*}
and 
\begin{align*}
    H(\kappa) : = \int_{\kappa}^{\infty} \frac{dt}{\sqrt{2 \pi}} \exp \left(- \frac{1}{2}t^{2} \right).
\end{align*}
Using the solution of these Eqs.(\ref{eq:extr_chitilde}) and (\ref{eq:extr_ptilde}), we express the stability condition of the successful solution as
\begin{align}
    \alpha > (1 - 2 \rho) H \left(\frac{1-\tilde{p}}{\sqrt{\tilde{\chi}}} \right) + \rho.
    \label{eq:linear_stability}
\end{align}
We can also obtain an analytical expression for the typical mean squared error value (MSE) value. 
The MSE is expressed as
\begin{align}
    [\text{MSE}]_{A, \bm{x}^{0}} = \rho - 2m + Q. \label{eq:MSE_replica}
\end{align}
Figure \ref{fig:figure1} (a)-(c) show the analytical results of this section for each $\gamma=\{0, 2, 10\}$.
The red dotted lines indicate the reconstruction limits of the signals estimated based on $L_{1}$-norm minimization, representing $\alpha_{c} = (1 - 2 \rho) H \left((1-\tilde{p})/\sqrt{\tilde{\chi}} \right) + \rho$.
The values of each pixel in Figure \ref{fig:figure1} show the mean squared error (MSE) as expressed by Eq.(\ref{eq:MSE_replica}).
$\{\tilde{Q}, \tilde{m}, \tilde{\chi}, \tilde{p}, Q, m, \chi, p\}$ are obtained by solving the saddle-point equations Eqs.(\ref{eq:extr_Qtilde})-(\ref{eq:extr_p}) iteratively.
Compression ratio $\alpha$ and non-zero elements rate $\rho$ are varied in steps of 0.01 from 0.01 to 1. Therefore, the heatmap consists of 100 $\times$ 100 = 10,000 pixels.
Our results in Figure \ref{fig:figure1} indicate that in binary compressed sensing, the larger the bias in the elements of the observation matrix, the wider the reconstruction success region.
This result is consistent with previous studies \cite{tanaka2018Performance2018IEEEInt.Symp.Inf.TheoryISIT, keiper2017CompressedLinearAlgebraanditsApplications,flinth2019RecoveryIEEETrans.Inform.Theory}.
\begin{figure}[htbp]
    \centering
    \includegraphics[width=0.9\linewidth]{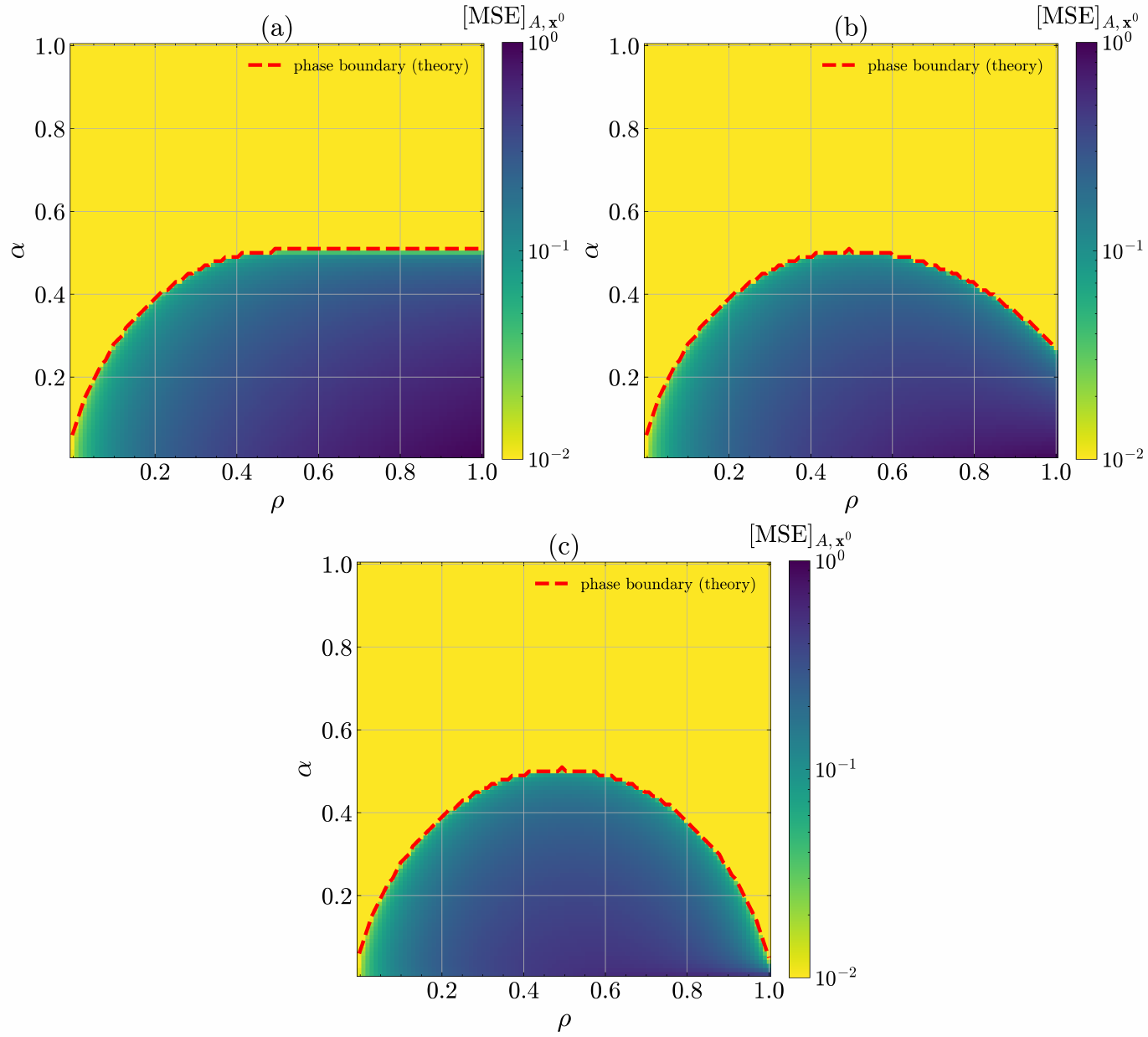}
    \caption{(a), (b), and (c) represent the phase diagrams of the analytical results for $\gamma=0 \,(\text{zero mean}), \gamma=2$, and $ \gamma=10$ cases, respectively. The vertical axis in each phase diagram represents the compression ratio $\alpha$, and the horizontal axis represents the non-zero elements rate $\rho$. The red dotted lines indicate the reconstruction limits of the signals estimated based on $L_{1}$-norm minimization, represented by $\alpha_{c} = (1 - 2 \rho) H \left((1-\tilde{p})/\sqrt{\tilde{\chi}} \right) + \rho$. Each pixel value shows the mean squared error (MSE) as expressed by Eq.(\ref{eq:MSE_replica}).}
    \label{fig:figure1}
\end{figure}
\newpage
\section{Numerical Simulations}
In this section, we verify the analytical results presented in the previous section through numerical simulations.
As mentioned in Section 1, the approximate message passing (AMP) algorithm generally becomes unstable when using observation matrices with non-zero means.
Therefore, in this section, we perform numerical simulations based on $L_{1}$-norm minimization using the interior point method implemented in Gurobi Optimizer, the commercial mathematical optimization solver.
The problem setup is the same as described in Section 2. We set the problem size to $N=500$ and take the configurational averages of 20 instances for each $\gamma=\{0, 2, 10\}$.
The results of the numerical simulations are shown in Figure \ref{fig:figure2} (a)-(c). The values of each pixel in Figure \ref{fig:figure2} represent the mean squared error (MSE) between the estimated and original signals $\|\hat{\bm{x}} - \bm{x}^{0}\|^{2}_{2}/N$.
Compression ratio $\alpha$ and non-zero elements rate $\rho$ are varied in steps of 0.01 in the range from 0.01 to 1.
The red dotted lines indicate the analytical reconstruction limit shown in Section 3.
Although several regions appear to be affected by the finite size effect, we can confirm that our analytical results are almost consistent with the numerical simulations.
\begin{figure}[htbp]
    \centering
    \includegraphics[width=0.9\linewidth]{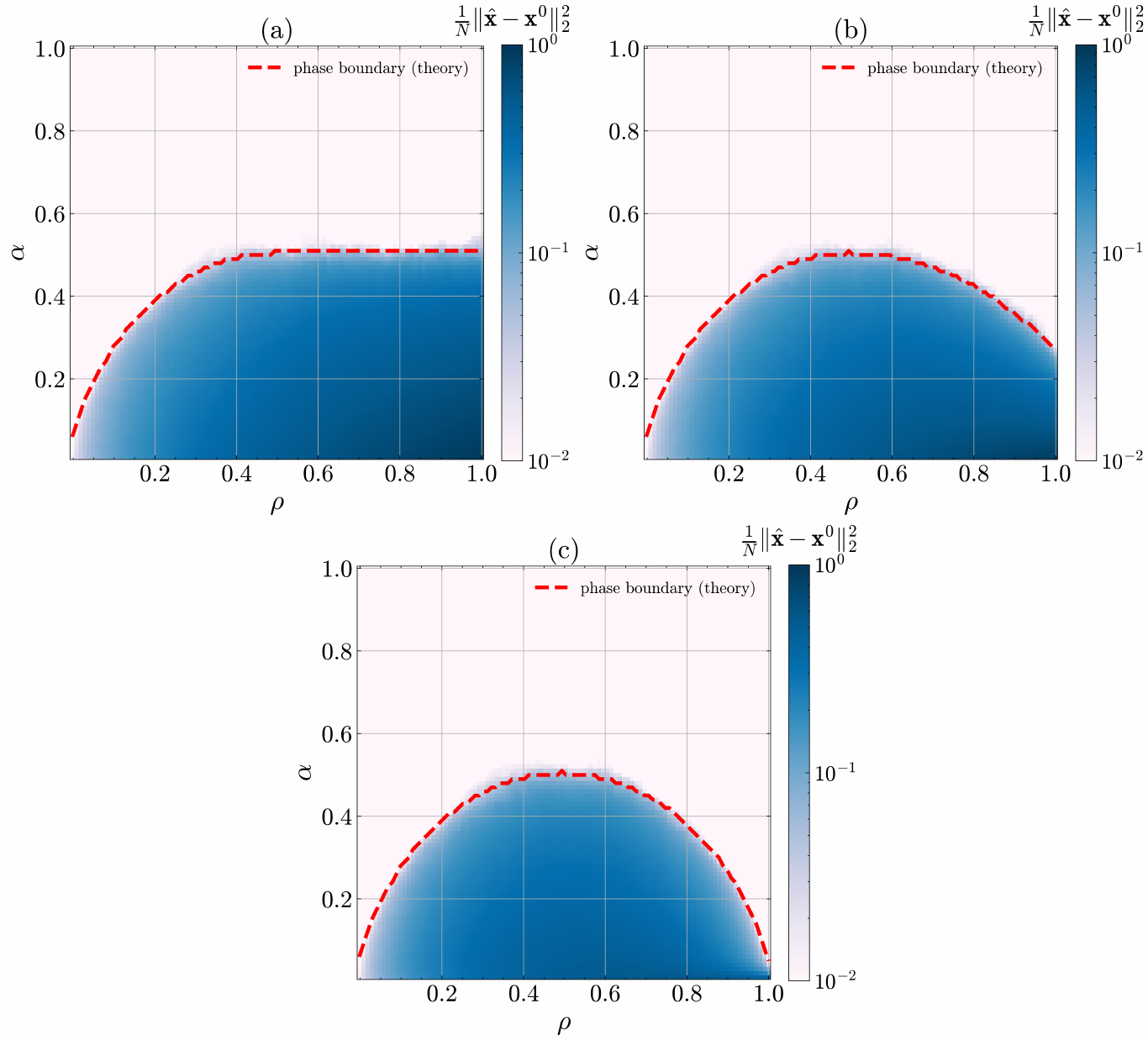}
    \caption{(a), (b), and (c) represent the phase diagrams of the numerical simulations results for $\gamma=0 \,(\text{zero mean}), \gamma=2$, and $ \gamma=10$ cases, respectively. Each pixel value shows the mean squared error (MSE) between the estimated and original signals $\|\hat{\bm{x}} - \bm{x}^{0}\|^{2}_{2}/N$. The red dotted lines indicate the analytical reconstruction limit shown in Section 3.
    }
    \label{fig:figure2}
\end{figure}
\newpage
\section{Conclusion}
In this paper, we focused on binary compressed sensing within the field of compressed sensing, where the elements of the original signal are limited to binary values $\{0, 1\}$, and we discussed typical evaluation based on $L_{1}$-norm minimization.
Through replica analysis considering general observation matrices, we showed that the reconstruction success region expands when each element of the observation matrix follows a biased probability distribution.
In addition, we verified our analysis through numerical simulations.
Our results are reasonable as they are consistent with several previous studies. 

There are two interesting future research topics.
The first concerns the application of approximate message passing (AMP) algorithms to observation matrices with non-zero means. 
As mentioned, the AMP algorithm can exhibit instability when applied to observation matrices with non-zero means.
Although modifications have been proposed to make AMP applicable beyond zero mean cases, as discussed in prior research \cite{tanaka2018Performance2018IEEEInt.Symp.Inf.TheoryISIT}.
However, it is unlikely that any AMP-like algorithm can achieve our theoretical phase boundaries.
A deeper understanding of the characteristics and behaviors of AMP-like message-passing algorithms remains a future work.
The second topic is related to the optimization of nonconvex regularization functions.
As mentioned, the typical performance in compressed sensing depends on the construction of the observation matrix and the choice of regularization functions.
In this paper, we used the $L_{1}$-norm, the convex regularization function, in the relaxed problem.
However, several previous studies on compressed sensing have shown that using nonconvex regularization functions allows for reconstructing the original signal from fewer observations than using the $L_{1}$-norm. \cite{sakata2021PerfectJ.Stat.Mech., nagano2023PhasePhys.Rev.E}.
Therefore, using nonconvex regularization functions in the relaxed problem of binary compressed sensing is an exciting setting to analyze in the future.

\begin{acknowledgement}
The authors thank the fruitful discussions with Manaka Okuyama.
This work is supported by JSPS KAKENHI Grant No. 23H01432.
Our study receives financial support from the programs for Bridging the gap between R\&D and the IDeal society (society 5.0) and Generating Economic and social value (BRIDGE) and the Cross-ministerial Strategic Innovation Promotion Program (SIP) from the Cabinet Office.
\end{acknowledgement}

\newpage
\section*{Appendix: Detailed derivation of Eq.(\ref{eq:extr_free_energy_density})}
\renewcommand{\theequation}{A-\arabic{equation}}
\setcounter{equation}{0}
This appendix presents the detailed derivation of the free energy density in Eq.(\ref{eq:extr_free_energy_density}).
First, we express the explicit form of $\left[Z_{\beta}^{n}(A, \bm{y})\right]_{A, \bm{x}^{0}}$ in Eq.(\ref{eq:replicated_system}), and then proceed with the subsequent analyses.
With variable transformation and order parameters as shown in Eqs.(\ref{eq:u_transformed})-(\ref{eq:order_params_q}), $\left[Z_{\beta}^{n}(A, \bm{y})\right]_{A, \bm{x}^{0}}$ can be written as follows.
\begin{align}
    \left[Z_{\beta}^{n}(A, \bm{y})\right]_{A, \bm{x}^{0}}
    =
    \prod_{a}
    \int_{0}^{1} d \bm{x}^{a} 
    \int d \bm{x}^{0}
    \exp(-\beta \|\bm{x}^{a}\|_{1})
    P_{0}(\bm{x}^{0})
    \prod_{\mu=1}^{M}
    \left[
        \prod_{a=1}^{n}
        \delta(u_{\mu}^{a}) 
    \right]_{\xi_{a}, z} 
    \nonumber \\
    \times
    \prod_{a, b}
    \int{dp} \int{dm} \int{dQ} \int{dq} \,
    \delta \left(p - \frac{1}{N} \sum_{i=1}^{N}x_{i}^{a} \right)
    \delta \left(m - \frac{1}{N}{\bm{x}^{0}}^{T} \bm{x}^{a} \right)
    \delta \left(Q - \frac{1}{N}{\bm{x}^{a}}^{T} \bm{x}^{a} \right)
    \delta \left(q - \frac{1}{N}{\bm{x}^{a}}^{T} \bm{x}^{b} \right) 
\end{align}
where $P_{0}(\bm{x}^{0}) = \prod_{i=1}^{N} \rho \delta(x_{i}^{0} - 1) + (1 - \rho)\delta(x_{i}^{0})$.
Furthermore, we introduce the following equations using the conjugate variables $\tilde{p}, \tilde{m}, \tilde{Q}, \tilde{q}$ which are introduced for the Fourier integral representation of the delta function.
\begin{align}
    \delta \left(p - \frac{1}{N} \sum_{i=1}^{N}x_{i}^{a} \right)
    & =
    \int d \tilde{p} \exp\left(- \tilde{p}\left(Np - \sum_{i=1}^{N}x_{i}^{a} \right) \right) \label{eq:delta_integral_p}
    \\
    \delta \left(m - \frac{1}{N}{\bm{x}^{0}}^{T} \bm{x}^{a} \right)
    & =
    \int d \tilde{m} \exp\left(- \tilde{m}\left(Nm - {\bm{x}^{0}}^{T} \bm{x}^{a} \right) \right) 
    \label{eq:delta_integral_m}
    \\
    \delta \left(Q - \frac{1}{N}{\bm{x}^{a}}^{T} \bm{x}^{a} \right)
    & = 
    \int d \tilde{Q} \exp\left(\frac{\tilde{Q}}{2}\left(NQ - {\bm{x}^{a}}^{T} \bm{x}^{a} \right) \right) 
    \label{eq:delta_integral_Q}
    \\
    \delta \left(q - \frac{1}{N}{\bm{x}^{a}}^{T} \bm{x}^{b} \right) 
    & =
    \int d \tilde{q} \exp\left(- \frac{\tilde{q}}{2}\left(Nq - {\bm{x}^{a}}^{T} \bm{x}^{b} \right) \right) 
    \label{eq:delta_integral_q}
\end{align}
Furthermore, by applying the Hubbard-Stratonovich transformation Eq.(\ref{eq:habast}) to Eq.(\ref{eq:delta_integral_q}), and carrying out the calculations for the part of $\prod_{\mu=1}^{M} \left[\prod_{a=1}^{n} \delta(u_{\mu}^{a}) \right]_{\xi_{a}, z}$, $\left[Z_{\beta}^{n}(A, \bm{y})\right]_{A, \bm{x}^{0}}$ can be expressed as shown in Eq.(\ref{eq:replicated_system_explicit})
\begin{align}
    \prod_{a\neq b}\exp\left(\frac{\tilde{q}}{2}{\bm{x}^{a}}^{T} \bm{x}^{b} \right)
    & =
    \int Dt \, 
    \prod_{a = 1}^{n}
    \prod_{i=1}^{N}
    \exp\left(\sqrt{\tilde{q}} x^{a}_{i} t - \frac{\tilde{q}}{2} (x^{a}_{i})^{2}\right) \label{eq:habast}
\end{align}
\begin{align}
    \left[Z_{\beta}^{n}(A, \bm{y})\right]_{A, \bm{x}^{0}} 
    =
    \int dp \int dm \int dQ \int dq 
    \int d \tilde{p} \int d\tilde{m} \int d\tilde{Q} \int d\tilde{q} \nonumber
    \\
    \exp
    \left(nN 
    \left(
        - \frac{\alpha[(\rho - 2m + q) + \gamma^{2}(\rho - p)^{2}]}{2(Q - q)} 
        - \frac{\alpha}{2} \log(2\pi)
        - p \tilde{p} - m \tilde{m} + \frac{1}{2}Q \tilde{Q} + \frac{1}{2}q \tilde{q} 
    \right) 
    \right) 
    \nonumber \\ 
    \times
        \left[
        \int Dt \,
            \prod_{a=1}^{n}
            \prod_{i=1}^{N}
            \left(
                \int_{0}^{1} dx^{a}_{i}
                \exp 
                \left(
                    - \frac{1}{2}(\tilde{Q} + \tilde{q})(x^{a}_{i})^{2} 
                    + (\tilde{m}x^{0}_{i} + \sqrt{\tilde{q}} t + \tilde{p})x^{a}_{i} 
                    - \beta \|x^{a}_{i} \|_{1}
                \right)
            \right)
        \right]_{x^{0}_{i}} \label{eq:replicated_system_explicit}
\end{align}
We showed the explicit form of $\left[Z_{\beta}^{n}(A, \bm{y})\right]_{A, \bm{x}^{0}}$ in Eq.(\ref{eq:replicated_system_explicit}).
However, Eq.(\ref{eq:replicated_system_explicit}) still has many integrals.
Here, by considering the saddle-point approximation $\int dx \exp(Ng(x)) \approx \exp(Ng(x^{*}))$ for the integrals with respect to $\{p, m, Q, q, \tilde{p}, \tilde{m}, \tilde{Q}, \tilde{\chi}\}$. Eq.(\ref{eq:replicated_system_explicit}) can be rewritten as:
\begin{align}
    \left[Z_{\beta}^{n}(A, \bm{y})\right]_{A, \bm{x}^{0}} 
    \approx
    \exp\left(nN 
    \left(
        - \frac{\alpha[(\rho - 2m + q) + \gamma^{2}(\rho - p)^{2}]}{2(Q - q)} 
        - p \tilde{p} - m \tilde{m} + \frac{1}{2}Q \tilde{Q} + \frac{1}{2}q \tilde{q} 
    \right) 
    \right) 
    \nonumber \\
    \times
        \left[
            \int Dt \,
            \exp 
            \left(nN
            \log
            \left(
                \int_{0}^{1}dx
                \exp 
                \left(
                    - \frac{1}{2}(\tilde{Q} + \tilde{q})x^{2} 
                    + (\tilde{m}x^{0} + \sqrt{\tilde{q}} t + \tilde{p})x 
                    - \beta \|x \|_{1}
                \right)
            \right)
            \right)
        \right]_{x^{0}} 
    \label{eq:replicated_system_explicit2}
\end{align}
where the parameters at the saddle points are denoted $p^{*} \rightarrow p, m^{*} \rightarrow m, Q^{*} \rightarrow Q, q^{*} \rightarrow q$, $\tilde{p}^{*} \rightarrow \tilde{p}, \tilde{m}^{*} \rightarrow \tilde{m}, \tilde{Q}^{*} \rightarrow \tilde{Q}, \tilde{q}^{*} \rightarrow \tilde{q}$.
The replica number $a$ and the vector index $i$ are omitted in Eq.(\ref{eq:replicated_system_explicit2}).
Next, let us consider the effect of the temperature on the parameters.
We assume that $Q - q \rightarrow \chi/\beta, \tilde{p} \rightarrow \beta \tilde{p}, \tilde{m} \rightarrow \beta \tilde{m}, \tilde{Q} + \tilde{q} \rightarrow \beta \tilde{Q}, \tilde{q} \rightarrow \beta^{2} \tilde{\chi}$, $\beta \rightarrow \infty$.
From these, we obtain the following expression for a part of Eq.(\ref{eq:replicated_system_explicit2})
\begin{align*}
        &\left[
            \int Dt
            \exp 
            \left(nN
            \log
            \left(
                \int_{0}^{1}dx
                \exp 
                \left(
                    - \frac{1}{2}(\tilde{Q} + \tilde{q})x^{2} 
                    + (\tilde{m}x^{0} + \sqrt{\tilde{q}} t + \tilde{p})x 
                    - \beta \|x \|_{1}
                \right)
            \right)
            \right)
        \right]_{x^{0}}
        \\
        & \rightarrow
        \left[
            \int Dt
            \exp 
            \left(nN
            \log
            \left(
                \int_{0}^{1}dx
                \exp
                \left(
                \beta
                    \left(
                        - \frac{1}{2}\tilde{Q}x^{2} 
                        + (\tilde{m}x^{0} + \sqrt{\tilde{\chi}} t + \tilde{p})x 
                        - \|x \|_{1}
                    \right)
                \right)
            \right)
            \right)
        \right]_{x^{0}} 
        \\
        & \approx
        \exp
        \left(
            nN
            \left[
            \int Dt \,
                \log
                \left(
                    \int_{0}^{1}dx
                    \exp
                    \left(
                    \beta
                    \left(
                        - \frac{1}{2}\tilde{Q}x^{2} 
                        + (\tilde{m}x^{0} + \sqrt{\tilde{\chi}} t + \tilde{p})x 
                        - \|x \|_{1}
                    \right)
                    \right)
                \right)
            \right]_{x^{0}}
        \right) \,\,\, (\because n \rightarrow 0)
        \\
        & \approx
        \exp
        \left(
            nN\beta
            \left[
                \int Dt \,
                \max_{\{0 \leq x \leq 1\}} 
                \left\{
                    -\frac{1}{2} \tilde{Q}x^{2} 
                    + \left(\tilde{m} x^{0} + \sqrt{\tilde{\chi}} t + \tilde{p} \right)x
                    - \|x\|_{1}
                \right\}
            \right]_{x^{0}}
        \right) \,\,\, (\because \beta \rightarrow \infty)
        \\
        & =
        \exp
        \left(
            - nN\beta
            \left(
                \rho \int Dt \, 
                \Psi (t; \tilde{Q}, \tilde{m}, \tilde{\chi}, \tilde{p}) 
                + (1 - \rho) \int Dt \, \Psi (t; \tilde{Q}, 0, \tilde{\chi}, \tilde{p})
            \right)
        \right)
\end{align*}
From these, $\left[Z_{\beta}^{n}(A, \bm{y})\right]_{A, \bm{x}^{0}}$ is expressed as Eq.(\ref{eq:replicated_system_explicit3}), and by substituting it into Eq.(\ref{eq:free_energy_density}), Eq.(\ref{eq:extr_free_energy_density}) is derived.
\begin{align}
    \left[Z_{\beta}^{n}(A, \bm{y})\right]_{A, \bm{x}^{0}} 
    &\approx 
    \exp 
    \left(
        nN\beta 
        \left[
            \frac{1}{2}Q \tilde{Q} 
            - m \tilde{m}
            - \frac{1}{2}\chi \tilde{\chi}
            - p \tilde{p}
            - \frac{\alpha[(\rho - 2m + Q) + \gamma^{2}(\rho - p)^{2}]}{2 \chi} \right. \right. \nonumber \\
            & \qquad \left. \left.
            - \rho \int Dt \, \Psi (t; \tilde{Q}, \tilde{m}, \tilde{\chi}, \tilde{p}) 
            - (1 - \rho) \int Dt \, \Psi (t; \tilde{Q}, 0, \tilde{\chi}, \tilde{p})
        \right] 
    \right) \label{eq:replicated_system_explicit3}
\end{align}

\bibliographystyle{jpsj}
\bibliography{main}
\end{document}